\documentclass{npw}
\usepackage{graphicx}

\usepackage[T1]{fontenc}
\usepackage[utf8]{inputenc}
\usepackage{cite,amsmath,amssymb,colordvi}

\begin{document}

\title{Finite element method for stochastic extended KdV equations}

\author{
  Anna Karczewska\email{a.karczewska@wmie.uz.zgora.pl}, 
 Maciej Szczeci\'nski \\
  \it Faculty of Mathematics, Computer Science and Econometrics \\
  Piotr Rozmej\email{p.rozmej@if.uz.zgora.pl} ~and Bartosz Boguniewicz \\
  \it Faculty of Physics and Astronomy, Institute of Physics\\
 \it  University of Zielona G\'ora, Szafrana 4a, 65-246 Zielona G\'ora, Poland
}

\normalsize \date{\today}

\maketitle

\begin{abstract}
The finite element method is applied to obtain numerical solutions to the recently derived nonlinear equation for shallow water wave problem  for several cases of bottom shapes. Results for time evolution of KdV solitons and cnoidal waves under stochastic forces are presented. Though small effects originating from second order dynamics may be obscured by stochastic forces, the main waves, both cnoidal and solitary ones, remain very robust against any distortions. 
\end{abstract}


\section{Introduction}\label{Int}

The Korteveg -- de Vries (KdV) equation appears as a model for the propagation of weakly nonlinear dispersive waves in several fields like gravity driven waves on a surface of  an incompressible irrotational inviscid fluid \cite{Whit,EIGR,Rem,DrJ,MS}, ion acoustic waves in plasma  \cite{EIGR}, impulse propagation in electric circuits \cite{Rem} and so on.

In hydrodynamical context the KdV equation is obtained as first order approximation of Euler equations with appropriate boundary conditions for the surface and the flat bottom. Small parameters, like $\alpha= a/H$, ratio of the wave amplitude $a$ to the fluid depth $H$ and $\beta=(H/L)^2$, where $L$ is a mean wavelength, are introduced.
In the derivation of KdV the velocity potential of the fluid is expanded as a  series with respect to the vertical coordinate and then only 
terms up to first order in small parameters are retained in all equations describing the system. 
Despite of its simplicity, KdV equation appeared to be a good approximation for many phenomena, see e.g.\ \cite{Whit,EIGR,Rem,DrJ} and focused enormous attention of physicists, mathematicians and engineers.

In order to extend possible applications of this kind of theories to the cases when parameters $\alpha,\beta$ are not very small the second order approximation of velocity potential was considered and subsequently second order KdV type equation (sometimes called extended KdV) was derived, see, e.g.\ \cite{MS,BS}. This equation, written in scaled coordinates and fixed reference frame has the following form \linebreak ($\eta(x,t)$ is the surface wave function)
\begin{eqnarray} \label{etaab}
\eta_t &+&\eta_x + \alpha\, \frac{3}{2}\eta\eta_x 
+\beta\,\frac{1}{6} \eta_{3x} - \frac{3}{8} \alpha^2 \eta^2\eta_x\\
&+&
 \alpha\beta\left(\frac{23}{24}\eta_x\eta_{2x}+\frac{5}{12}\eta\eta_{3x} \right)+\frac{19}{360}\beta^2\eta_{5x} = 0 \nonumber 
\end{eqnarray}
Here and in the following the indexes denote partial derivatives, that is $\eta_x\equiv \frac{\partial \eta}{\partial x}, u_t\equiv \frac{\partial \eta}{\partial t}$ and so on.

There were also attempts to extend the KdV theory to cases when the bottom of the fluid container is not flat. 
This subject is extremely important to understand the behaviour of waves coming to shallower regions and such phenomenon as tsunami creation.
However, until last year there were no studies in this direction which could directly incorporate terms from the bottom function into the wave equation.
In \cite{KRR,KRI}, some of us derived second order KdV type equation containing terms directly related to the bottom function $h(x)$ 
\begin{eqnarray} \label{etaabd}
\eta_t \!\!&\!\!+\!\!&\!\!
\eta_x + \alpha\, \frac{3}{2}\eta\eta_x +\beta\,\frac{1}{6} \eta_{3x} \\
\!\!&\!\!-\!\!&\!\!
\frac{3}{8} \alpha^2 \eta^2\eta_x 
+ \alpha\beta \!\left(\!\frac{23}{24}\eta_x\eta_{2x}\!+\!\frac{5}{12}\eta\eta_{3x}\! \right)\!+\!\frac{19}{360}\beta^2\eta_{5x} \nonumber \\
\!\!&\!\!+\!\!&\!\!
\beta\delta\left(-\frac{1}{2\beta}(h\eta)_x +\frac{1}{4} \left(h_{2x}\eta\right)_x -\frac{1}{4} \left(h\eta_{2x}\right)_x\right) =0,
\nonumber
\end{eqnarray}
where $\delta=a_h/H$ is another small parameter, $a_h$ stands for the amplitude of the bottom variations and $H$ is the undisturbed (mean) depth of the fluid. 
 It is assumed that $h(x)$ is continuous, at least three times  differentiable  and 
$\displaystyle \lim_{x\to\pm\infty} h(x)=0$.  
For details of derivation of that equation, see \cite{KRR,KRI}.
Note that when the bottom is flat, $\delta=0$, the equation (\ref{etaabd}) reduces to (\ref{etaab}). 

The present paper aims to find numerical solutions to stochastic versions of second order KdV type equations  (\ref{etaab}) and (\ref{etaabd}). In order to do this we first try to follow the finite element method (FEM) used by Debussche and Printems in \cite{DebP}.  Their method was good enough for the the  stochastic Korteweg-de Vries equation of the form \cite[Eq.~1.2]{DebP}
\begin{equation} \label{kdv}
u_t+u\,u_x+\epsilon\, u_{xxx}= \gamma\, \dot{\xi}.
\end{equation}
In Eq.~(\ref{kdv}) the noise term $\xi(x,t)$ is a Gaussian process with correlations 
\begin{equation} \label{chidot}
\mathbb{E}\,\dot{\xi}(x,t)\,\dot{\xi}(y,s) = c(x-y)\,\delta(t-s)
\end{equation}
and $\gamma$ is the amplitude of the noise.
Equation (\ref{kdv}) with rhs equal zero is the Korteweg-de Vries equation written in a moving reference system with coordinates scaled in a particular way. This form was convenient for the authors in order to apply the finite element method (FEM) in their numerical simulation.

In the case of periodic boundary conditions the noise term $\xi$ has to be introduced in a different way. Since the Brownian motions are nowhere differentiable we have to introduce the mathematical form of  (\ref{kdv}).

Let $(\Omega,\mathcal{F},(\mathcal{F}_t)_{t\ge 0},\mathbb{P})$ denote a stochastic basis.
The It\^o form of (\ref{kdv}) can be written in the form
\begin{equation} \label{kdvI}
d u +\left(u \frac{\partial u}{\partial x} + \epsilon \frac{\partial^3 u}{\partial x^3} \right) dt= \gamma\, \Phi\, dW.
\end{equation}
In (\ref{kdvI}),  $W$ is a Wiener process on $L^2(0,L)$ of the form
\begin{equation} \label{Wien}
W(t,x) = \sum_{i=0}^\infty \beta_i(t)\, e_i(x),
\end{equation}
where $\{e_i \}_{i\in\mathbb{N}}$ is an orthonormal basis of $L^2(0,L)$ and $\{\beta_i \}_{i\in\mathbb{N}}$  is a sequence of independent real valued Brownian motions, defined on the stochastic basis. In (\ref{kdvI}), $\Phi$ is an appropriate linear map from  $L^2(0,L)$ to 
 $L^2(0,L)$. For more details, see \cite{DebP}.

\section{Numerical approach} \label{nsym}


Our aim was to extend the approach used in \cite{DebP}  in order to numerically solve second order stochastic version of the equation with an uneven bottom (\ref{etaabd})
\begin{eqnarray} \label{etaabds}
\eta_t \!\!&\!\!+\!\!&\!\!
\eta_x + \alpha\, \frac{3}{2}\eta\eta_x +\beta\,\frac{1}{6} \eta_{3x} \\
\!\!&\!\!-\!\!&\!\!
\frac{3}{8} \alpha^2 \eta^2\eta_x 
+ \alpha\beta \!\left(\!\frac{23}{24}\eta_x\eta_{2x}\!+\!\frac{5}{12}\eta\eta_{3x}\! \right)\!+\!\frac{19}{360}\beta^2\eta_{5x} \nonumber \\
\!\!&\!\!+\!\!&\!\!
\beta\delta\left(-\frac{1}{2\beta}(h\eta)_x +\frac{1}{4} \left(h_{2x}\eta\right)_x -\frac{1}{4} \left(h\eta_{2x}\right)_x\right) = \gamma\, \dot{\xi}.
\nonumber
\end{eqnarray}
Note that this equation, in contrast to KdV equation, has to be solved in the fixed coordinate system because transformation to a moving frame would make the bottom function time dependent.

Setting $\delta=0$ in (\ref{etaabds}) one obtains second order stochastic KdV type equation (that is the equation for the flat bottom)
\begin{eqnarray} \label{etaabs}
\eta_t &+&\eta_x + \alpha\, \frac{3}{2}\eta\eta_x 
+\beta\,\frac{1}{6} \eta_{3x} - \frac{3}{8} \alpha^2 \eta^2\eta_x\\
&+&
 \alpha\beta\left(\frac{23}{24}\eta_x\eta_{2x}+\frac{5}{12}\eta\eta_{3x} \right)+\frac{19}{360}\beta^2\eta_{5x} =  \gamma\, \dot{\xi} \nonumber .
\end{eqnarray}
which can be solved within the same algorithm.

The details of numerical scheme for solution of equations (\ref{etaab}), (\ref{etaabd})
were described in \cite{KRSB}. Therefore in this paper we only briefly summarize that description emphasizing the stochastic part. We focus on (\ref{etaabds}) because in our scheme (\ref{etaabs}) is the particular case of  (\ref{etaabds}) when $\delta=0$.

We adapt Crank-Nicholson scheme for time evolution using time step $\tau$.
Introducing the following variables: 
\begin{equation} \label{notat}
v=\eta_x,~~ w=v_x,~~ p=w_x,~~  q=p_x,~~ g=h_{xx} 
\end{equation} 
we can write fifth order differential equation (\ref{etaabds}) as the coupled set of first order differential equations 
\begin{eqnarray} \label{CrABD}
\eta^{n+1}- \eta^{n}-\Phi \left( W^{n+1} -W^n \right) \hspace{10ex}\nonumber \\
+\tau \frac{\partial }{\partial x} \left[ \eta^{n+\frac{1}{2}} +\frac{3\alpha}{4} 
\left(\eta^{n + \frac{1}{2}}\right)^{2} 
\!\! + \! \frac{\beta}{6} w^{n + \frac{1}{2}} \!  \right. \nonumber \\ \hspace{-3ex}
-\!\frac{1}{8}\alpha^2 \left(\eta^{n+\frac{1}{2}}\right)^3\!\!
+ \alpha\beta\left(\!\frac{13}{48}\left(v^{n + \frac{1}{2}}\right)^2  \right. \nonumber
\\ \left. \hspace{-3ex}
\!+\!\frac{5}{12}\left(\eta^{n + \frac{1}{2}}w^{n + \frac{1}{2}\!}\right)\! \right)\! 
+ \frac{19}{360}\beta^2 \left(q^{n + \frac{1}{2}}\right)  \nonumber 
\\  
\frac{1}{4}\beta\delta  \left(-\frac{2}{\beta}\left(h^{n + \frac{1}{2}}\eta^{n + \frac{1}{2}\!}\right)  \right. \hspace{5ex}  \\ \left. \left.
+\eta^{n + \frac{1}{2}}g^{n + \frac{1}{2}} + h^{n + \frac{1}{2}}w^{n + \frac{1}{2}} \right) \right] 
\!\!&\!\!=\!\!&\!\! 0,  \nonumber \\
\frac{\partial }{\partial x}\eta^{n + \frac{1}{2}}-v^{n + \frac{1}{2}} \!\!&\!\!=\!\!&\!\! 0,  \nonumber\\
\frac{\partial }{\partial x} v^{n + \frac{1}{2}}-w^{n + \frac{1}{2}} \!\!&\!\!=\!\!&\!\!  0,  \nonumber \\
\frac{\partial }{\partial x} w^{n + \frac{1}{2}}-p^{n + \frac{1}{2}} \!\!&\!\!=\!\!&\!\! 0,  \nonumber\\ \frac{\partial }{\partial x} p^{n + \frac{1}{2}}-q^{n + \frac{1}{2}} \!\!&\!\!=\!\!&\!\! 0,  \nonumber 
\end{eqnarray} 
where 
\begin{equation} \label{e12a}\def\arraystretch{1.2} 
\begin{array}{rclcrcl}
\eta^{n+\frac{1}{2}}\!\!&\!\!=\!\!&\!\!\frac{1}{2}\left( \eta^{n+1} + \eta^{n} \right),  
\!\!&\!\! \!\!&\!\! v^{n+\frac{1}{2}}\!\!&\!\!=\!\!&\!\!\frac{1}{2}\left( v^{n+1} + v^{n} \right),  \\
w^{n+\frac{1}{2}}\!\!&\!\!=\!\!&\!\!\frac{1}{2}\left( w^{n+1} + w^{n} \right), \!\!&\!\! \!\!&\!\!
p^{n+\frac{1}{2}}\!\!&\!\!=\!\!&\!\!\frac{1}{2}\left( p^{n+1} + p^{n} \right),  \\
q^{n+\frac{1}{2}}\!\!&\!\!=\!\!&\!\!\frac{1}{2}\left( q^{n+1} + q^{n} \right), \!\!&\!\!
 \!\!&\!\!
h^{n+\frac{1}{2}}\!\!&\!\!=\!\!&\!\!\frac{1}{2}\left( h^{n+1} + h^{n} \right),  \\
g^{n+\frac{1}{2}}\!\!&\!\!=\!\!&\!\!\frac{1}{2}\left( g^{n+1} + g^{n} \right).  \!\!&\!\!\!\!&\!\! \!\!&\!\!\!\!&\!\!
\end{array}
\end{equation}
In (\ref{CrABD}), $\eta^{n}\!=\!\eta(x,n\tau), \eta^{n+1\!}=\!\eta(x,(n+1)\tau)$ and so on.

\subsection{Finite element method}

Since solutions to stochastic equation are not expected to be smooth, we follow the arguments given in \cite{DebP}  and apply Petrov-Galerkin space discretization and the finite element method. We use a piecewise linear shape function and piecewise constant test functions. We consider wave motion on the interval $x\in [0,L]$ with periodic boundary conditions. Let $N\in \mathbb{N}$, then we use a mesh $M_{\chi}$ of points $x_j= j\chi$, $j=0,1,\ldots,N$, where $\chi =L/N$. Let $V^1_{\chi}$ be a space of a picewise linear functions $\varphi(x)$, such that $\varphi(0)=\varphi(L)$, defined as
\begin{equation} \label{phi}
\varphi_{j}(x) = \left\{ \begin{array}{lll}
\frac{1}{\chi}(x-x_{j-1}) & \mbox{if} &  x \in [x_{j-1},x_{j}] \\
\frac{1}{\chi}(x_{j+1}-x) & \mbox{if} &  x \in [x_{j},x_{j+1}] \\
0 &  & \mbox{otherwise} . \end{array} \right. 
\end{equation}
For test functions we choose the space of piecewise constant functions $\psi(x)\in V^0_{\chi}$, where 
\begin{equation} \label{psi}
\psi_{j}(x) = \left\{ \begin{array}{lll}
1 & \mbox{if} &  x \in [x_{j},x_{j+1})\\
0 &  &\mbox{otherwise}. \end{array} \right.
\end{equation}

Approximate solution and its derivatives may be expanded in the basis 
(\ref{phi}) 
\begin{equation}  \label{etfi}
\hspace{-1.5ex}\begin{array}{lcl} 
\eta_{\chi}^{n}(x) = \displaystyle \sum_{j=1}^{N}a_j^n\,\varphi_j(x), \!&\! &\!
 v_{\chi}^{n}(x) =  \displaystyle\sum_{j=1}^{N}b_j^n\,\varphi_j(x),\\
 w_{\chi}^{n}(x) = \displaystyle \sum_{j=1}^{N}c_j^n\,\varphi_j(x), \!&\! &\!
 p_{\chi}^{n}(x) =  \displaystyle\sum_{j=1}^{N}d_j^n\,\varphi_j(x),\\
 q_{\chi}^{n}(x) =  \displaystyle\sum_{j=1}^{N}e_j^n\,\varphi_j(x). \!&\! &\!
\end{array} \hspace{-1ex}
\end{equation}
Therefore in a weak formulation we can write (\ref{CrABD}) as
\begin{eqnarray} \label{CrABw}
\left(\eta_{\chi}^{n+1}- \eta_{\chi}^{n},\psi_i\right)  - \left(\Phi \left( W_{\chi}^{n+1} -W_{\chi}^{n} \right),\psi_i \right)  \hspace{7ex}   \nonumber \\
+ \tau \left\{   \left(\partial_x\eta_{\chi}^{n+\frac{1}{2}},\psi_i\right) 
  +\frac{3\alpha}{4} \left(
\partial_x\left(\eta_{\chi}^{n + \frac{1}{2}}\right)^{2},\psi_i\right)  \right. \hspace{3ex}
\nonumber \\
 +  \frac{\beta}{6}\left(\partial_x w_{\chi}^{n + \frac{1}{2}},\psi_i\right)    
-\frac{1}{8}\alpha^2 \left(\partial_x\left(\eta_{\chi}^{n+\frac{1}{2}}\right)^3,\psi_i\right)\nonumber  \\
+ \alpha\beta\left[\frac{13}{48}\left(\partial_x\left(v_{\chi}^{n + \frac{1}{2}}\right)^2,\psi_i\right)  \right. \\  \left.\hspace{-3ex}
+\frac{5}{12}\left(\partial_x\left(\eta_{\chi}^{n + \frac{1}{2}}w_{\chi}^{n + \frac{1}{2}}\right),\psi_i\right)    \right] \nonumber  \\
+ \frac{19}{360}\beta^2\left( \partial_x\left(q_{\chi}^{n + \frac{1}{2}}\right),\psi_i\right)   \nonumber  \\
+ \frac{1}{4}\beta\delta  \left(\partial_x\left[-\frac{2}{\beta}\left(h^{n + \frac{1}{2}}\eta_{\chi}^{n + \frac{1}{2}}\right) \right. \right. \nonumber \\
 \left.\left. \left. +\eta_{\chi}^{n + \frac{1}{2}}g^{n + \frac{1}{2}} + h^{n + \frac{1}{2}}w_{\chi}^{n + \frac{1}{2}} \right],\psi_i\right)  \right\}
 \!\!&\!\!=\!\!&\!\!      0, \nonumber  
\end{eqnarray}
\begin{eqnarray*} 
\left(\partial_x\eta_{\chi}^{n + \frac{1}{2}},\psi_i\right)-\left(\!v_{\chi}^{n + \frac{1}{2}},\psi_i\!\right) \!\!&\!\!=\!\!&\!\! 0, \nonumber\\ 
\left(\partial_x v_{\chi}^{n + \frac{1}{2}},\psi_i\right)-\left(\!w_{\chi}^{n + \frac{1}{2}},\psi_i\!\right) \!\!&\!\!=\!\!&\!\!0, \nonumber \\
\left(\partial_x w_{\chi}^{n + \frac{1}{2}},\psi_i\right)-\left(\!p_{\chi}^{n + \frac{1}{2}},\psi_i\!\right) \!\!&\!\!=\!\!&\!\! 0,  \nonumber\\ 
\left(\partial_x p_{\chi}^{n + \frac{1}{2}},\psi_i\right)-\left(\!q_{\chi}^{n + \frac{1}{2}},\psi_i\!\right)  \!\!&\!\!=\!\!&\!\! 0, \nonumber 
\end{eqnarray*}
for any $i=1,\dots,N$, where abbreviation  $\partial_x$ is used for $\frac{\partial }{\partial x}$.
Here and in the following, $$(f,g):=\int_0^L f(x) g(x) dx$$ denotes the scalar product of functions.

In order to obtain a noise in space, at each time step $n$ and each node $j$ a random number $\kappa^{\chi,\tau}_{j,n}$ is computed according to a normal law and such that it forms a sequence of independent random variables. Then we can set 
\begin{eqnarray*}
 \Phi \left( W_{\chi}^{n+1} -W_{\chi}^{n} \right) &=& \sqrt{\tau}\, \sum_{j=1}^{N} \frac{1}{|| \phi_j ||}_{L^2(0,L)} \kappa^{\chi,\tau}_{j,n}\, \phi_j \\
&=& \sqrt{\tau}\, N_\phi  \sum_{j=1}^{N} \kappa^{\chi,\tau}_{j,n}\, \phi_j,
\end{eqnarray*}
where notation $ \displaystyle N_\phi\:= \frac{1}{|| \phi_j ||}_{L^2(0,L)}$ was introduced for abbreviation.

Insertion (\ref{etfi}) into (\ref{CrABw}) yields a system of coupled linear equations for coefficients $a_j^n, b_j^n,  c_j^n, d_j^n, e_j^n$. Solution of this system supplies an approximate solution of (\ref{etaab}) given in the mesh points $x_j$.

Denote 
\begin{equation} \label{defC}\def\arraystretch{1.4}
\begin{array}{l}
 C^{(1)}_{ij} := (\varphi_{j},\psi_{i}), \\
 C^{(2)}_{ij} := (\varphi'_{j},\psi_{i}), \\
 C^{(3)}_{ijk}:= (\varphi'_{j}\varphi_{k}+\varphi_{j}\varphi'_{k},\psi_{i}),\\ C^{(4)}_{ijkl}:= \left(\left[\varphi'_{j}\varphi_{k}\varphi_{l} + \varphi_{j}\varphi'_{k}\varphi_{l} + \varphi_{j}\varphi_{k}\varphi'_{l} \right],\psi_i \right)
\end{array}
\end{equation}  
where $\varphi'_j=\frac{d\varphi}{dx}(x_j)$. Simple integration shows that 
\begin{equation} \label{C1}\def\arraystretch{1.4}
C^{(1)}_{ij}=\left\{ \begin{array}{rl}
\frac{1}{2} \chi &\quad  \textrm{if} \quad i=j \quad \mbox{or} \quad i=j-1 \\
0 & \quad \textrm{otherwise},
\end{array} \right. 
\end{equation}  
\begin{equation} \label{C2}\def\arraystretch{1.4}
C^{(2)}_{ij}=\left\{ \begin{array}{rl}
-1&\quad  \textrm{if} \quad i=j \\
 1&\quad  \textrm{if} \quad i=j-1 \\
0 & \quad \textrm{otherwise}.
\end{array} \right. 
\end{equation}  
A little more complicated calculation yields
\begin{equation} \label{C3}\def\arraystretch{1.4}
C^{(3)}_{ijk}= C^{(2)}_{ij}\, \delta_{jk} \quad \mbox{and}  \quad
C^{(4)}_{ijkl} = C^{2}_{ij}\,\delta_{jk}\,\delta_{kl}.
\end{equation}  

Properties (\ref{C3}) allow to reduce double and triple sums arising in nonlinear terms in (\ref{CrABw}) to single ones. 

The final system of nonlinear equations for coefficients $a^{n+1}_{j},b^{n+1}_{j},c^{n+1}_{j},d^{n+1}_{j},e^{n+1}_{j}$ of expansion of the solution in the basis $\{\phi_i\}$ has the form (for details of derivation, see \cite{KRSB}) 

\begin{eqnarray} \label{matr3}
 \sum_{j=1}^{N}  \left\{ \left(a^{n+1}_{j} - a^{n}_{j} 
+  \sqrt{\tau}\, N_\phi \kappa^{\chi,\tau}_{j,n}\right) C^{(1)}_{ij} \right.\hspace{4ex} \nonumber \!\!&\!\! \!\!&\!\!  \\
+ \tau \left[ \frac{1}{2} (b^{n+1}_{j} + b^{n}_{j}) C^{(1)}_{ij} \right.
\hspace{16ex}
 \!\!&\!\! \!\!&\!\!  \\
+  \left( \alpha \frac{3}{16} (a^{n+1}_{j} + a^{n}_{j})^2  
+ \beta \frac{1}{12} (c^{n+1}_{j} + c^{n}_{j}) \right. \!\!&\!\! \!\!&\!\!  \nonumber \\
- \alpha^2 \frac{1}{64}  
(a^{n+1}_{j} + a^{n}_{j})^3   
+ \alpha\beta \frac{13}{192}  (b^{n+1}_{j} + b^{n}_{j})^2 \!\!&\!\! \!\!&\!\!\nonumber  \\
+\alpha\beta \frac{5}{96}  (a^{n+1}_{j} + a^{n}_{j}) (c^{n+1}_{j}  + c^{n}_{j})  \!\!&\!\! \!\!&\!\!\nonumber  \\  \left.  \left. \left.
+ \beta^2 \frac{19}{720}(e^{n+1}_{j}  + e^{n}_{j}) \right) C^{(2)}_{ij} \right]\right\}
\!\!&\!\!= \!\!&\!\! 0, \nonumber \\
\sum_{j=1}^{N} \left[(a^{n+1}_{j} + a^{n}_{j}) C^{(2)}_{ij} - (b^{n+1}_{j} + b^{n}_{j}) C^{(1)}_{ij} \right] \!\!&\!\!=\!\!&\!\!0, \nonumber\\
\sum_{j=1}^{N} \left[(b^{n+1}_{j} + b^{n}_{j}) C^{(2)}_{ij} - (c^{n+1}_{j} + c^{n}_{j}) C^{(1)}_{ij} \right] \!\!&\!\!=\!\!&\!\!0, \nonumber \\
\sum_{j=1}^{N} \left[(c^{n+1}_{j} + c^{n}_{j}) C^{(2)}_{ij} - (d^{n+1}_{j} + d^{n}_{j}) C^{(1)}_{ij} \right] \!\!&\!\!=\!\!&\!\!0, \nonumber\\
\sum_{j=1}^{N} \left[(b^{n+1}_{j} + b^{n}_{j}) C^{(2)}_{ij} - (e^{n+1}_{j} \!+\! e^{n}_{j}) C^{(1)}_{ij} \right] \!\!&\!\!=\!\!&\!\!0, \nonumber 
\end{eqnarray}
where  $i=1,2,\ldots,N$.

Define 5$N$-dimensional vector of expansion coefficients (\ref{etfi})
\begin{equation} \label{vec1}
X^n= \left(\!\begin{array}{c}A^n\\ B^n\\ C^n \\ D^n \\ E^n\end{array}\!\right),
\end{equation} 
where 
\begin{equation*} \label{vec2} 
A^n= \left(\!\!\begin{array}{c} a_1^n\\ a_2^n\\ \vdots \\ a_N^n \end{array} \!\!\right)\!,~
B^n= \left(\!\!\begin{array}{c} b_1^n\\ b_2^n\\ \vdots \\ b_N^n \end{array} \!\!\right)\!,~
C^n= \left(\!\!\begin{array}{c} c_1^n\\ c_2^n\\ \vdots \\ c_N^n \end{array}\!\! \right)\!,
\end{equation*}
\begin{equation*} \label{vec2a}
D^n= \left(\!\!\begin{array}{c} d_1^n\\ d_2^n\\ \vdots \\ d_N^n \end{array} \!\!\right)\!, \quad \mbox{and} \quad
E^n= \left(\!\!\begin{array}{c} e_1^n\\ e_2^n\\ \vdots \\ e_N^n \end{array}
\!\! \right)\!.
\end{equation*}

In (\ref{matr3}), $A^{n+1}, B^{n+1}, C^{n+1}, D^{n+1}, E^{n+1}$ represent the unknown coefficients whereas $A^{n}, B^{n}, C^{n}, D^{n}, E^{n}$ the known ones. Note that the system (\ref{matr3}) is a nonlinear one. 

In an abbreviated form the set (\ref{matr3}) can be written as 
\begin{equation} \label{mv1}
F_i(X^{n+1},X^{n}) =  0, \quad i=1,2,\ldots,5N.
\end{equation}
Since this equation is nonlinear we can use Newton method at each time step. That is, we find $ X^{n+1}$ by iterating the equation
\begin{equation} \label{iter}
(X^{n+1})_{m+1}=(X^{n+1})_{m} + J^{-1}(X^{n+1})_{m}  =  0,
\end{equation}
where $J^{-1}$ is the inverse of the Jacobian of the  $F(X^{n+1},X^{n})$   (\ref{mv1}). Choosing $(X^{n+1})_{0} =X^{n}$ we usually obtain the approximate solution to (\ref{mv1}), $(X^{n+1})_m$ in $m=3-5$ iterations with very good precision.
The Jacobian itself is a particular sparse matrix $(5N\times 5N)$ with the following block structure
\begin{equation} \label{J2}
J= \left( \begin{array}{ccccc} (A_a) & (A_b) &  (A_c) & (0) & (A_e) \\ (C2) & -(C1) &  (0) & (0) & (0)  \\ (0) & (C2) & -(C1) & (0) & (0) \\
(0) & (0) &(C2) & -(C1) & (0) \\  (0) & (0) & (0) &(C2) & -(C1)  
\\\end{array} \right),
\end{equation}
where each block $(\cdot )$ is a two-diagonal sparse $(N\times N)$ matrix. The matrix $A_a$ is given by
\begin{equation} \label{Aa}
A_a\!=\! \left(\! \!\begin{array}{ccccccc} 
a^{1}_{1} & 0 & 0 & \cdots & 0 & a^{1}_{N-1} & a^{1}_{N} \\
a^{2}_{1} &a^{2}_{2} & 0 & \cdots & 0 & 0 & a^{2}_{N} \\
0 & a^{3}_{2} &  a^{3}_{3} & 0 & \cdots & 0 & 0 \\
\vdots & \vdots & \vdots & \ddots & \vdots & \vdots & \vdots \\
0 & 0 & \cdots & a^{N-3}_{N-4} & a^{N-3}_{N-3} & 0 & 0 \\
0 & 0 & \cdots & 0 & a^{n-2}_{N-3} & a^{N-2}_{N-2} & 0 \\
a^{N}_{1} & 0 & \cdots & 0 & 0 & a^{N-1}_{N} & a^{n}_{N} \!
\end{array}\!\! \right)\!\!.
\end{equation}
In (\ref{Aa}) the nonzero elements of $(A_a)$ are given by 
\begin{equation} \label{Aai}
a^i_j=\frac{\partial\, F_i}{\partial\, a^{n+1}_j}, 
\end{equation}
where $ F_i, ~i=1,\ldots,N$ is given by the first equation of the set (\ref{matr3}).
Elements in the upper right and lower left corners come from periodic boundary conditions.
Matrices $(A_b), (A_c), (A_e)$ have the same structure as  $(A_a)$, only elements $a^i_j$ have to be replaced, respectively, by $$b^i_j=\frac{\partial\, F_i}{\partial\, b^{n+1}_j},\quad c^i_j=\frac{\partial\, F_i}{\partial\, c^{n+1}_j}\quad \mbox{and}\quad e^i_j=\frac{\partial\, F_i}{\partial\, e^{n+1}_j}.$$ 
Matrix $(A_d)$ vanishes since there is no fourth order derivative in the extended KdV equation and $d^n$ does not appear in $F$. 

Matrices $C1$ and $C2$ are constant. They are defined as $Ck$, ~$k=1,2$
\begin{equation} \label{Ck1}
Ck\!=\! \left(\! \!\begin{array}{ccccc} 
C^{(k)}_{11} & 0  & \cdots &  C^{(k)}_{11} & C^{(k)}_{1N} \\
C^{(k)}_{21} & C^{(k)}_{22}  & \cdots &  0 & C^{(k)}_{2N} \\
\vdots & \vdots  & \ddots &  \vdots & \vdots \\
0 & 0 & \cdots & C^{(k)}_{N-1N-1} & 0 \\
C^{(k)}_{N1} & 0 & \cdots  &  C^{(k)}_{N-1N} & C^{(k)}_{NN} \!
\end{array}\!\! \right)\!\! ,
\end{equation}
where $C^{(k)}_{ij}$ are defined in (\ref{C1}) and  (\ref{C2}).

\section{Results of simulations}\label{result}

In this section some results of numerical calculations of soliton waves with stochastic forces are presented. Simulations were performed by solving the set of equations (\ref{matr3}) step by step. The main aim was to compare time evolution of waves described by second order KdV-type equation with bottom dependent term with stochastic forces to those obtained without these forces (obtained in previous paper \cite{KRSB}).  In order to do this several cases of time evolution are presented in the following convention. For each particular case of the bottom function $h(x)$  a sequence of three figures is presented in which the amplitude of the stochastic force is 0, 0.001 and 0.002, respectively. In this way the influence of an increasing stochastic term on the wave evolution is exhibited. In all presented cases $\delta=0.2$, that is, the amplitude of the bottom variation is 20\% of the average water depth. In order to avoid overlaps of the wave profiles at different time instants, the consecutive profiles are shifted vertically by 0.15.

\subsection{Soliton waves} \label{sol}

\begin{figure}[tbh]  
\resizebox{0.99\columnwidth}{!}{\includegraphics{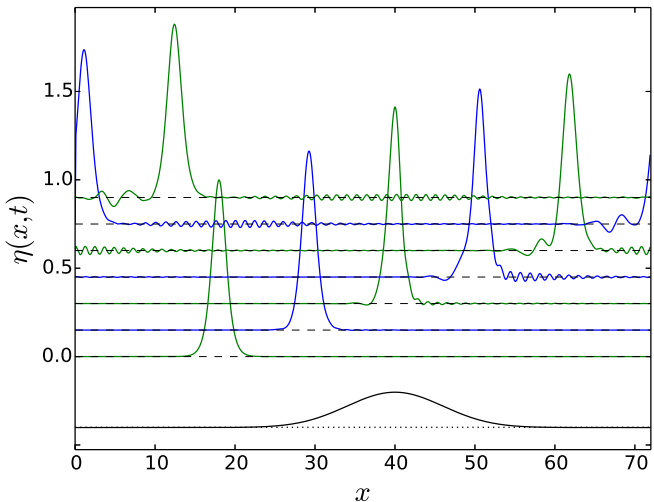}}
\vspace{-2mm}
\caption{
Time evolution of the initial KdV soliton governed by the extended KdV equation (\ref{etaabd}) obtained with FEM method, by numerical solution of the set of equations (\ref{matr3}) with $\gamma=0$. Dashed lines represent the undisturbed fluid surface.} 
 \label{g000}
\end{figure}

\begin{figure}[tbh]  
\resizebox{0.99\columnwidth}{!}{\includegraphics{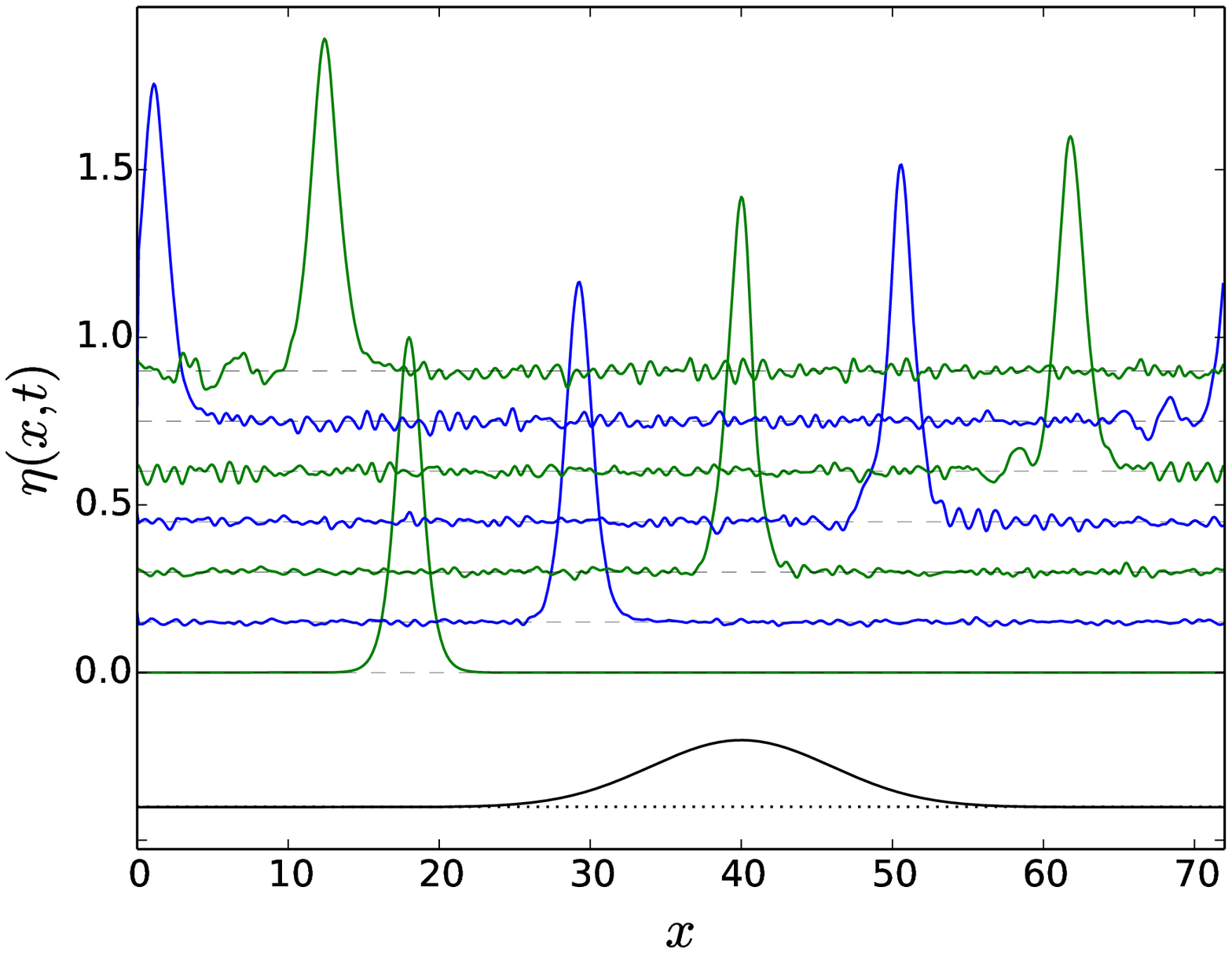}}
\vspace{-2mm}
\caption{
The same as in Fig.~\ref{g000} but with a moderate amplitude of stochastic force, $\gamma=0.001$.} 
 \label{g001}
\end{figure}

\begin{figure}[tbh]  
\resizebox{0.99\columnwidth}{!}{\includegraphics{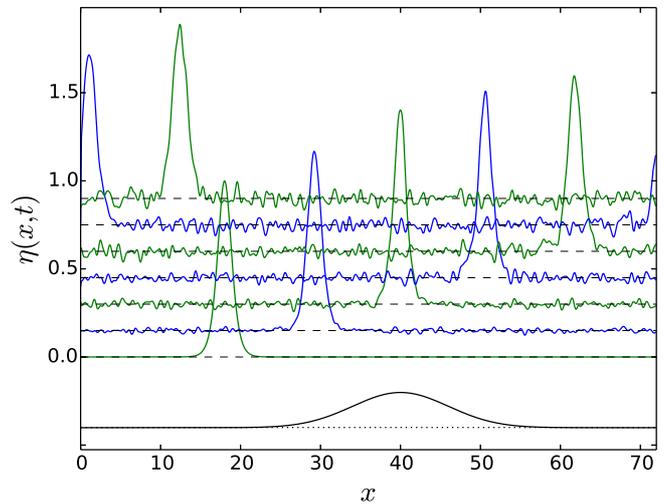}}
\vspace{-2mm}
\caption{The same as in Fig.~\ref{g000} but with a larger amplitude of stochastic force, $\gamma=0.002$.} 
 \label{g002}
\end{figure}

In figures \ref{g000}-\ref{g002} we compare time evolution of the wave (initialy a KdV soliton) when the bottom function represents a wide Gaussian hump, $$h_1(x)=\delta \exp\left(-\frac{(x - 40)^2}{72}\right). $$

\begin{figure}[tbh]  
\resizebox{0.99\columnwidth}{!}{\includegraphics{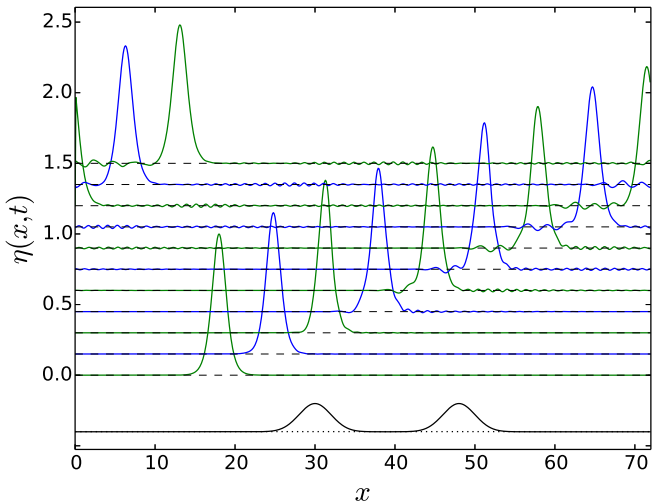}}
\vspace{-2mm}
\caption{The same as in Fig.~\ref{g000} but for a double Gaussian hump bottom function.} 
 \label{dg000}
\end{figure}

\begin{figure}[tbh]  
\resizebox{0.99\columnwidth}{!}{\includegraphics{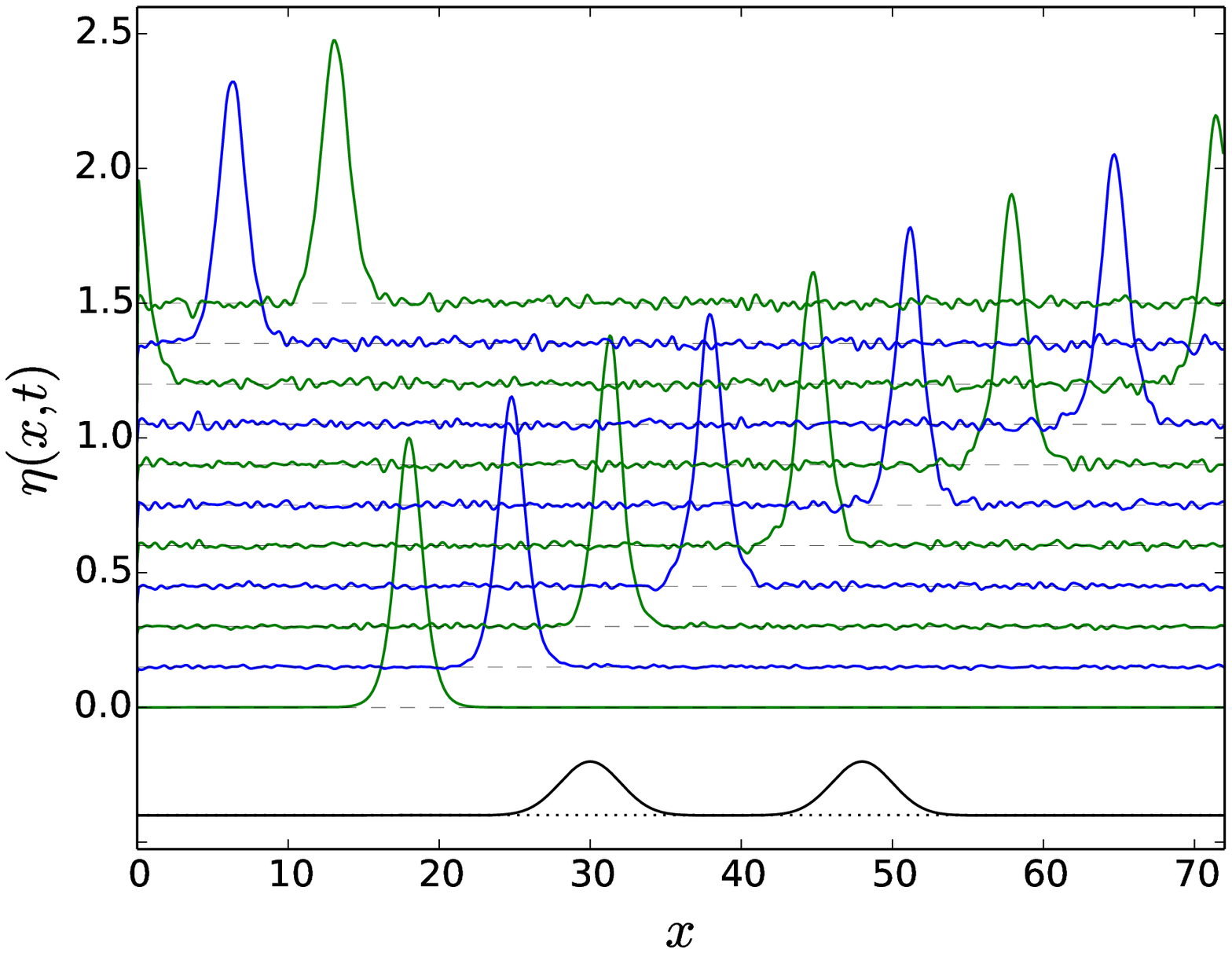}}
\vspace{-2mm}
\caption{The same as in Fig.~\ref{dg000} but with a moderate amplitude of stochastic force, $\gamma=0.001$.} 
 \label{dg001}
\end{figure}

\begin{figure}[tbh]  
\resizebox{0.99\columnwidth}{!}{\includegraphics{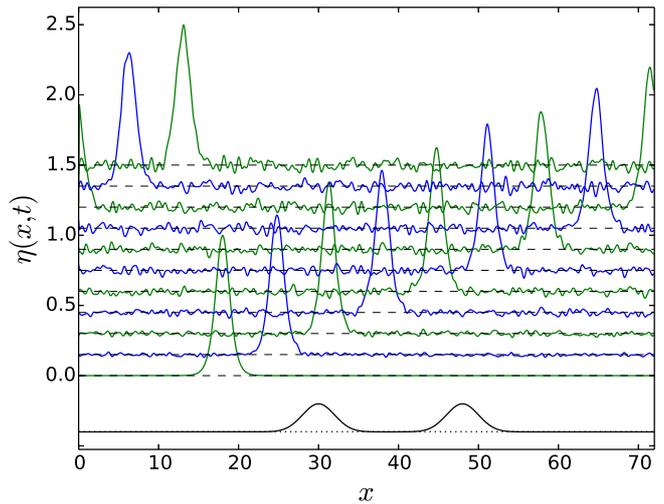}}
\vspace{-2mm}
\caption{The same as in Fig.~\ref{dg001} but with a larger amplitude of stochastic force, $\gamma=0.002$.} 
 \label{dg002}
\end{figure}

In figures \ref{dg000}-\ref{dg002} we compare time evolution of the wave (initialy a KdV soliton) when the bottom function represents a double Gaussian hump, $$h_2(x)=\delta \left[\exp\left(-\frac{(x - 30)^2}{4}\right) + \exp\left(-\frac{(x - 48)^2}{4}\right)\right].$$

\begin{figure}[tbh]  
\resizebox{0.99\columnwidth}{!}{\includegraphics{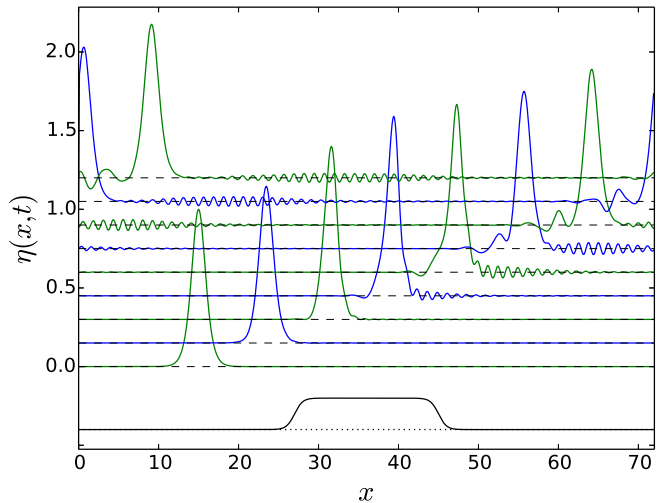}}
\vspace{-2mm}
\caption{The same as in Fig.~\ref{g000} but for the bottom function in the form of an extended hump.} 
 \label{h000}
\end{figure}

\begin{figure}[tbh]  
\resizebox{0.99\columnwidth}{!}{\includegraphics{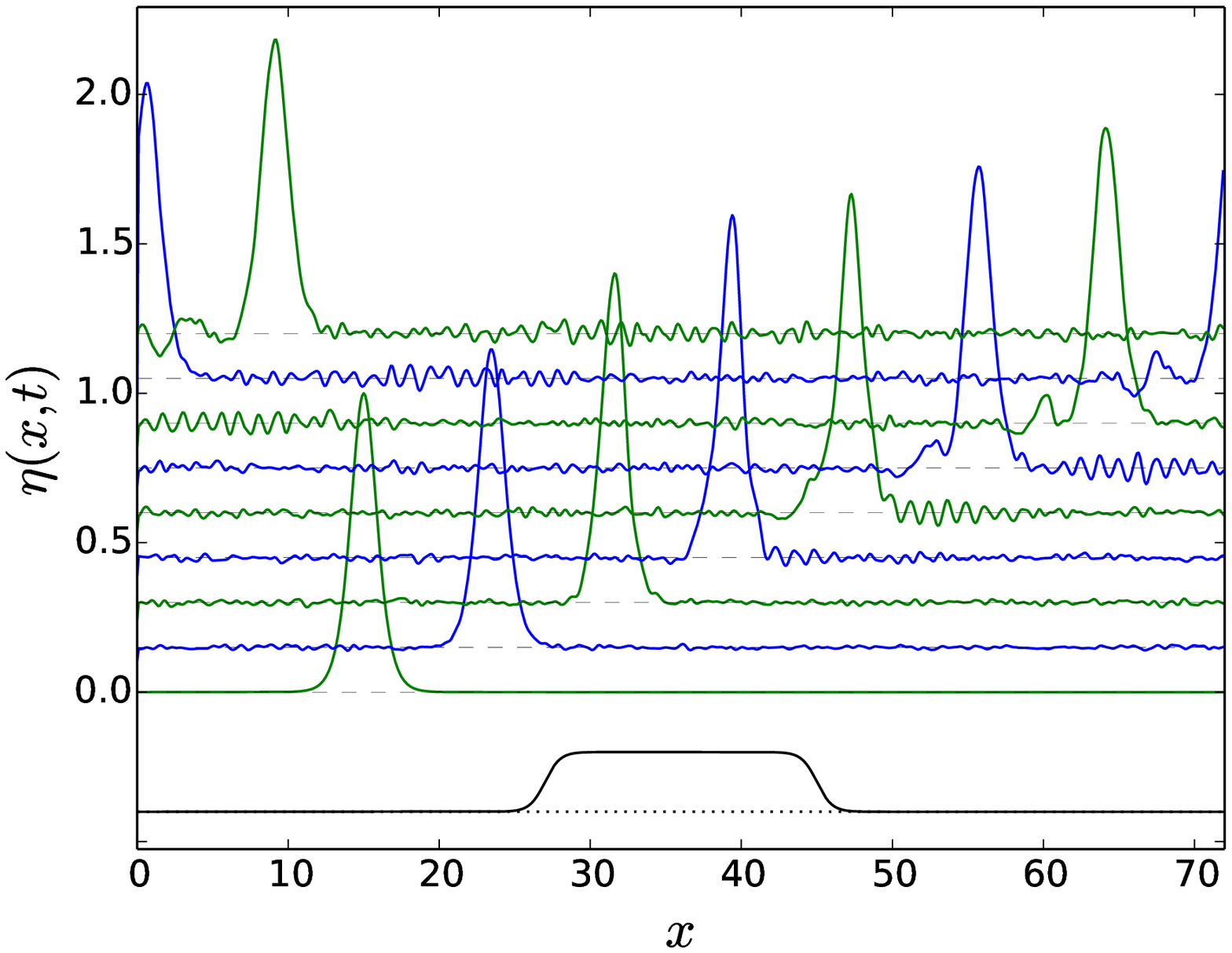}}
\vspace{-2mm}
\caption{The same as in Fig.~\ref{h000} but with a moderate amplitude of stochastic force, $\gamma=0.001$.} 
 \label{h001}
\end{figure}

\begin{figure}[tbh]  
\resizebox{0.99\columnwidth}{!}{\includegraphics{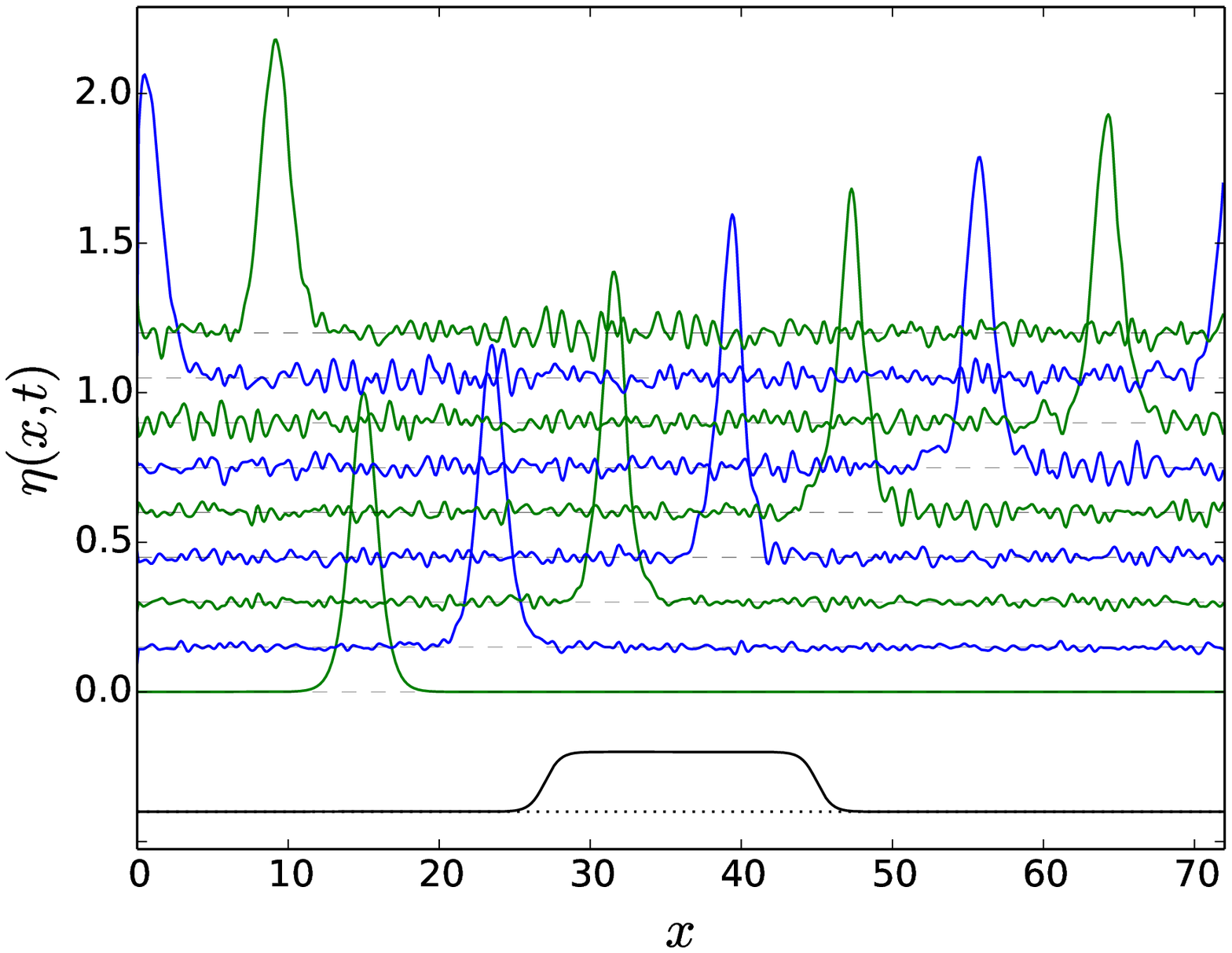}}
\vspace{-2mm}
\caption{The same as in Fig.~\ref{h001} but with a larger amplitude of stochastic force, $\gamma=0.002$.} 
 \label{h002}
\end{figure}

In figures \ref{h000}-\ref{h002} we compare time evolution of the initial  KdV soliton  when \,the bottom function represents an extended hump, $$h_3(x)=\delta \left(\frac{ \tanh(x - 27) - \tanh(x - 45)}{2} \right).$$

\begin{figure}[tbh]  
\resizebox{0.99\columnwidth}{!}{\includegraphics{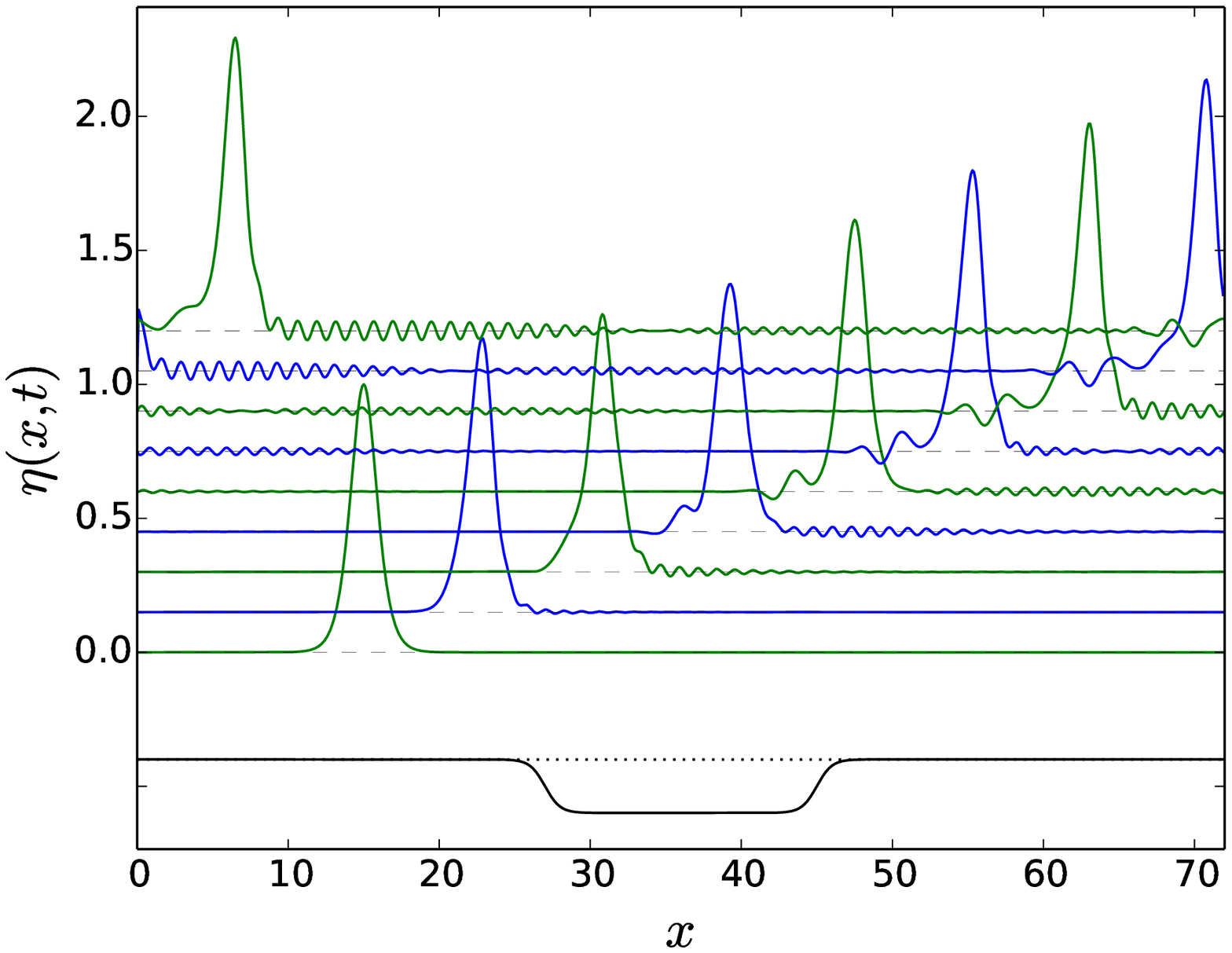}}
\vspace{-2mm}
\caption{The same as in Fig.~\ref{g000} but for the bottom function in the form of an extended valley.} 
 \label{v000}
\end{figure}

\begin{figure}[tbh]  
\resizebox{0.99\columnwidth}{!}{\includegraphics{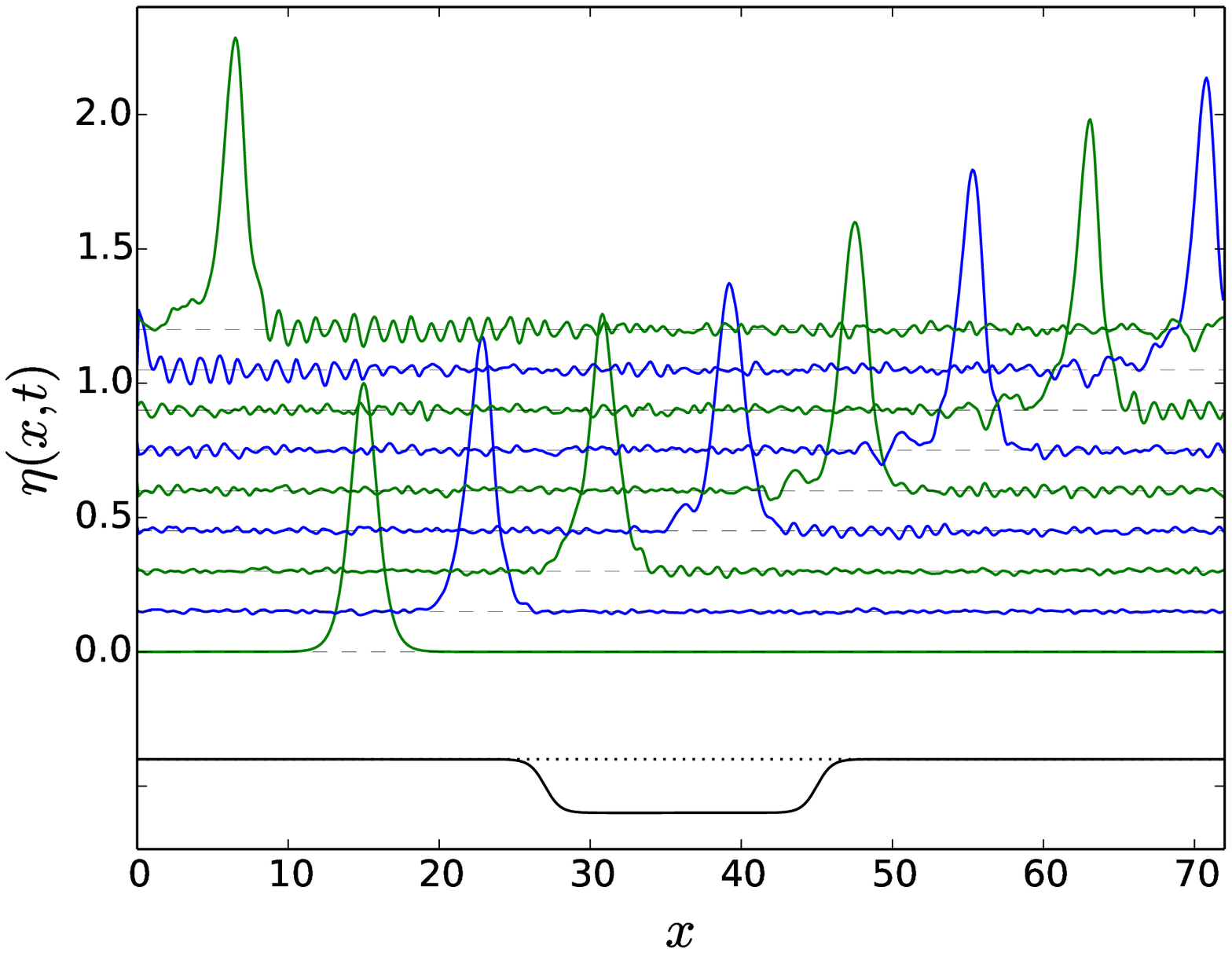}}
\vspace{-2mm}
\caption{The same as in Fig.~\ref{v000} but with a moderate amplitude of stochastic force, $\gamma=0.001$} 
 \label{v001}
\end{figure}

\begin{figure}[tbh]  
\resizebox{0.99\columnwidth}{!}{\includegraphics{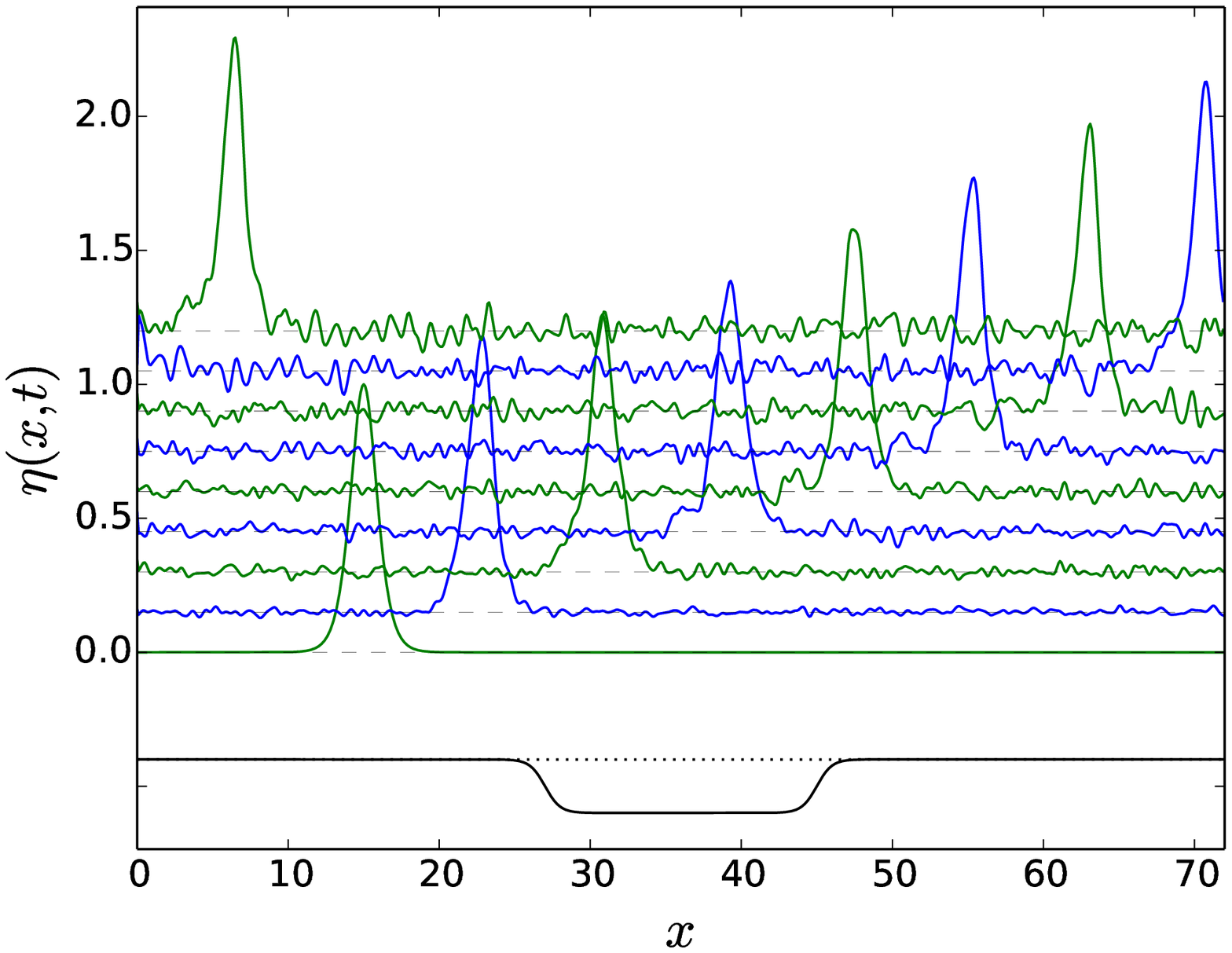}}
\vspace{-2mm}
\caption{The same as in Fig.~\ref{h001} but with a larger amplitude of stochastic force, $\gamma=0.002$.} 
 \label{v002}
\end{figure}

In figures \ref{v000}-\ref{v002} time evolution of the initial  KdV soliton is compared for different amplitude of the stochastic term when the bottom function represents a valley, $h_4(x)=-h_3(x)$. 

In all the examples presented, with different shapes of bottom functions, one observes the same general trend. When the amplitude of the stochastic force is relatively small  ($\gamma=0.001$), some small structures originated from second order terms in the evolution equation (\ref{etaabd}) can be still visible. Simultaneously, the main soliton wave is disturbed only a little by the stochastic term.

When the amplitude of the stochastic force increases, through $\gamma=0.0015$ which case is not shown here, to $\gamma=0.002$, the structures from second order terms begin to be more and more obscured by the noise and are not visible at $\gamma=0.002$. The main wave, however, appears to be strongly resistant to the noise and preserves its soliton character.

This character is preserved also for times much longer than in presented examples. To see that
we needed, however, to adapt our different code, based on finite difference method, to the  stochastic case. That code proved to be very efficient in numerical calculations presented in \cite{KRR,KRI,KRI2}. The reasons why the finite difference method is more effective than finite element method presented here are explained in detail in the next section.

\subsection{Cnoidal waves} \label{cno}

The cnoidal solutions to KdV are expressed by the Jacobi elliptic {\sf cn$^2$} function. 
The explicit expression for such a solution is the following, see, e.g.\ \cite{Ding}:
\begin{equation} \label{cnsol}
\eta(x,t) = \eta_2 + H \textrm{cn}^2\left(\left. \frac{x-ct}{\Delta} \right\vert m \right),
\end{equation}
where 
\begin{equation} \label{cnsol1}
\eta_2=\frac{H}{m}\left(1-m-\frac{E(m)}{K(m)} \right),\quad \Delta = h\sqrt{\frac{4 m h}{3 H}},
\end{equation}
and
\begin{equation} \label{cnsol2}
c=\sqrt{gh}\left[1+\frac{H}{mh} \left(1-\frac{m}{2} - \frac{3 E(m)}{2 K(m)}\right) \right].
\end{equation}
Solution (\ref{cnsol})-(\ref{cnsol2}) is written in dimensional quantities, where $H$ is the wave height, $h$ is mean water depth, $g$ is gravitational acceleration and $m$ is an elliptic parameter. $K(m)$ and $E(m)$ are complete elliptic integrals of the first and the second kind, respectively. The value of $m\in [0,1]$ governs the shape of the wave. When $m\to 1$ the solution tends to a soliton wave with distance between the peaks going to infinity. When $m\to 0$ the cnoidal solution tends to usual sinusoidal wave.\vspace{0.5ex}

For our equations (\ref{etaabd}) and ({\ref{etaab}}) we have to express the formulas (\ref{cnsol})-(\ref{cnsol2}) in dimensionless variables.

\begin{figure}[tbh]  
\resizebox{0.99\columnwidth}{!}{\includegraphics{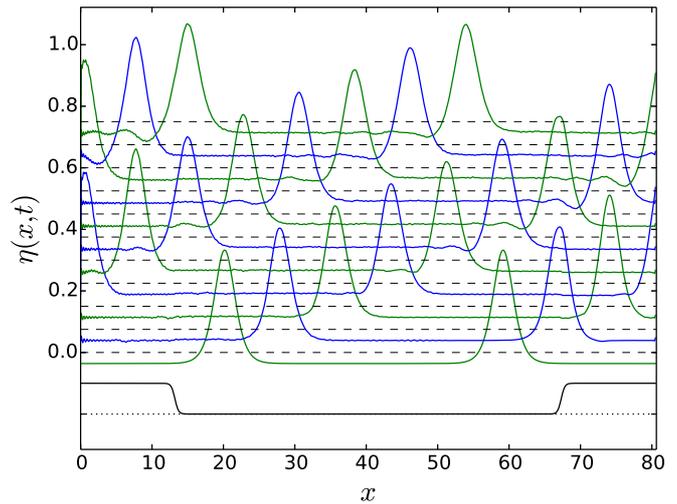}}
\vspace{-2mm}
\caption{Time evolution, according to eq.~(\ref{etaabd}), of the cnoidal wave for the bottom function in the form of an extended valley. The $x$ interval is equal to the double wavelength of the cnoidal wave with $m$=1-10$^{-8}$.} 
 \label{m8000}
\end{figure}

\begin{figure}[tbh]  
\resizebox{0.99\columnwidth}{!}{\includegraphics{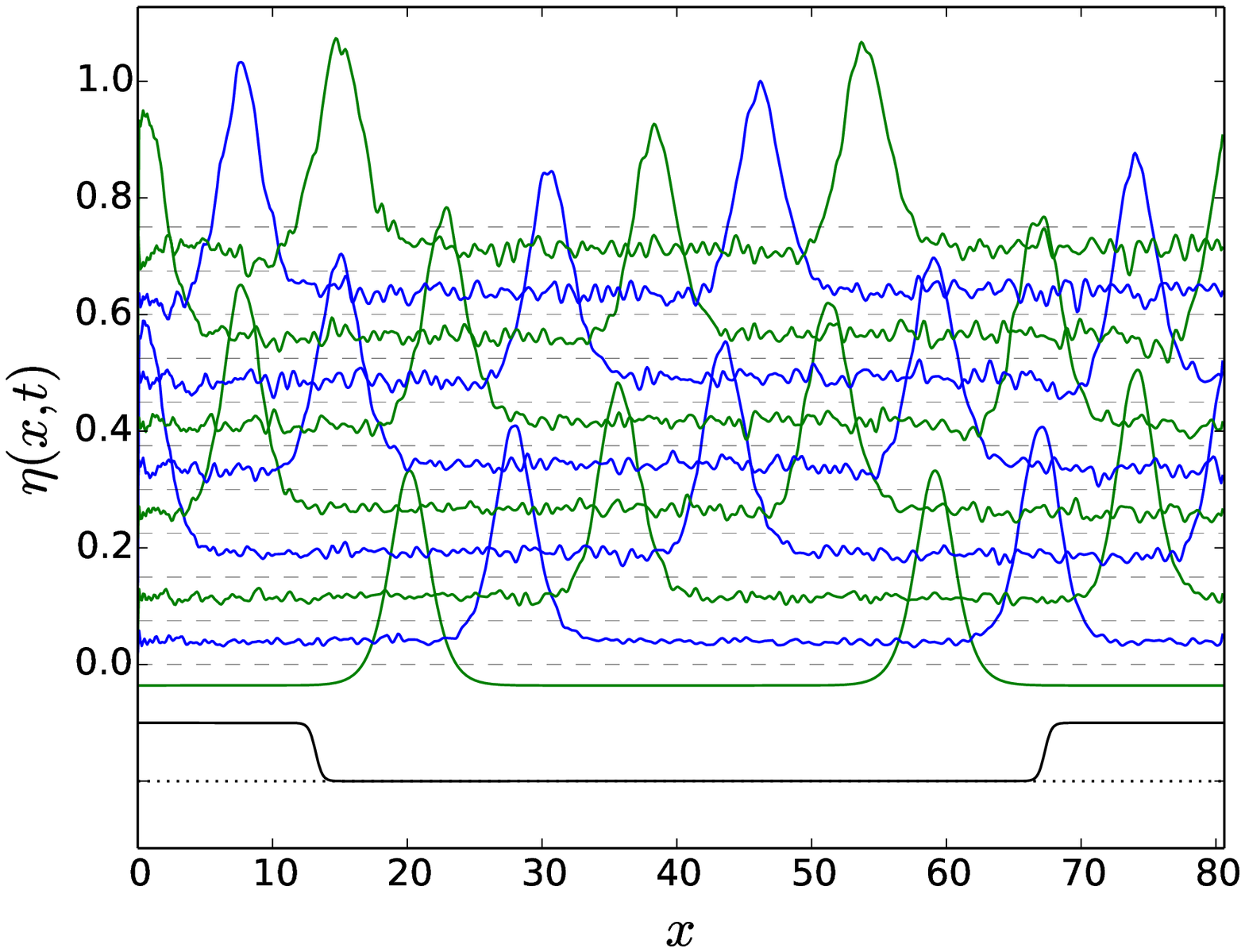}}
\vspace{-2mm}
\caption{The same as in Fig.~\ref{m8000} but with a moderate amplitude of stochastic force, $\gamma=0.001$} 
 \label{m8001}
\end{figure}

\begin{figure}[tbh]  
\resizebox{0.99\columnwidth}{!}{\includegraphics{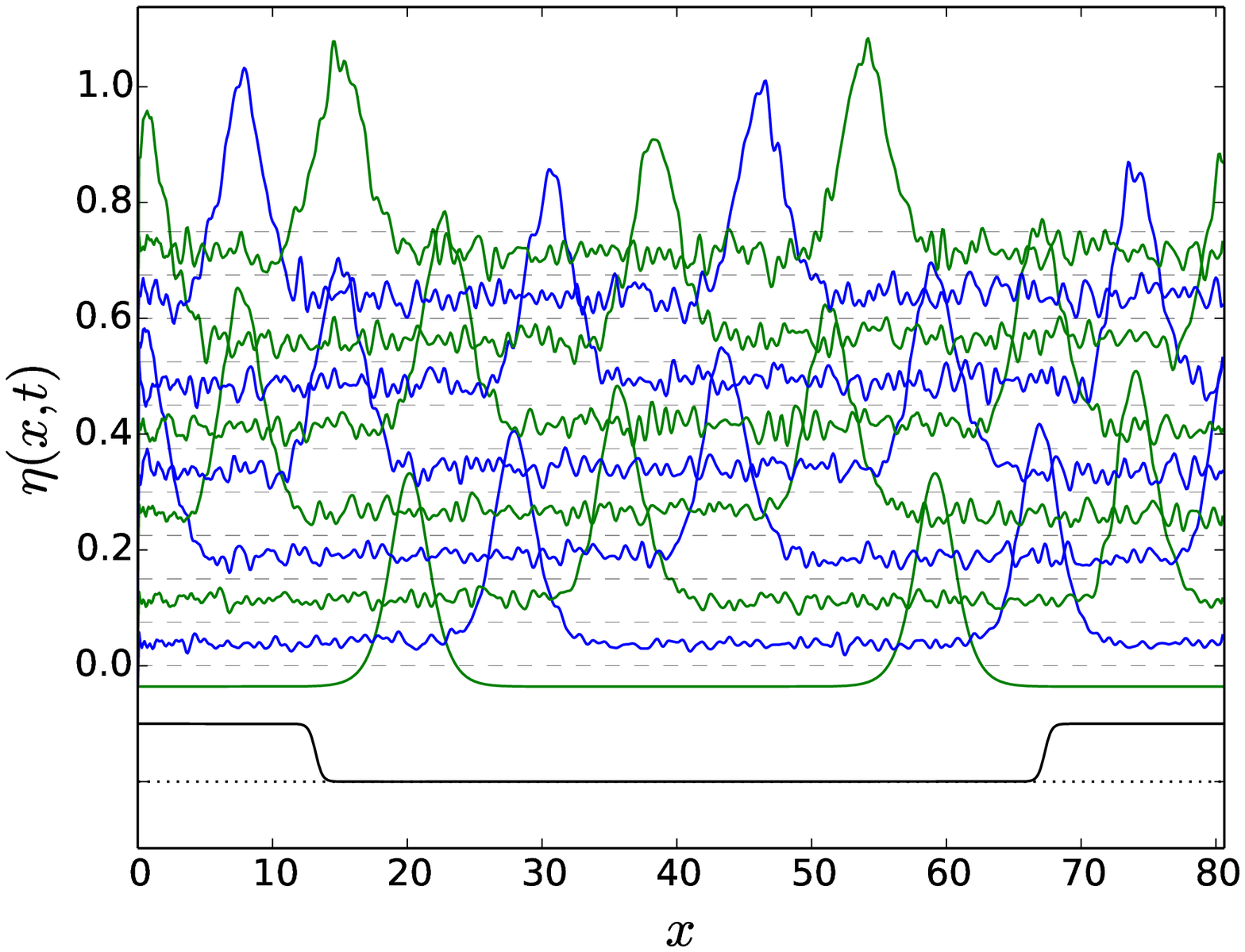}}
\vspace{-2mm}
\caption{The same as in Fig.~\ref{m8001} but with a larger amplitude of stochastic force, $\gamma=0.0015$.} 
 \label{m80015}
\end{figure}

In figures \ref{m8000}-\ref{m80015} we display time evolution of the wave, initialy cnoidal solution of KdV equation with $m=1-10^{-8}$ for $\gamma=0, 0.001$ and  $0.0015$.
In this case the bottom function is $h(x)=\delta \frac{1}{2}[-\textrm{tanh}(2(x-8.6)-\frac{1}{2})+\textrm{tanh}(2(x-66.5552)-\frac{1}{2})]$ and the wavelength of the cnoidal wave is $d\approx 40.324$.

\begin{figure}[tbh]  
\resizebox{0.99\columnwidth}{!}{\includegraphics{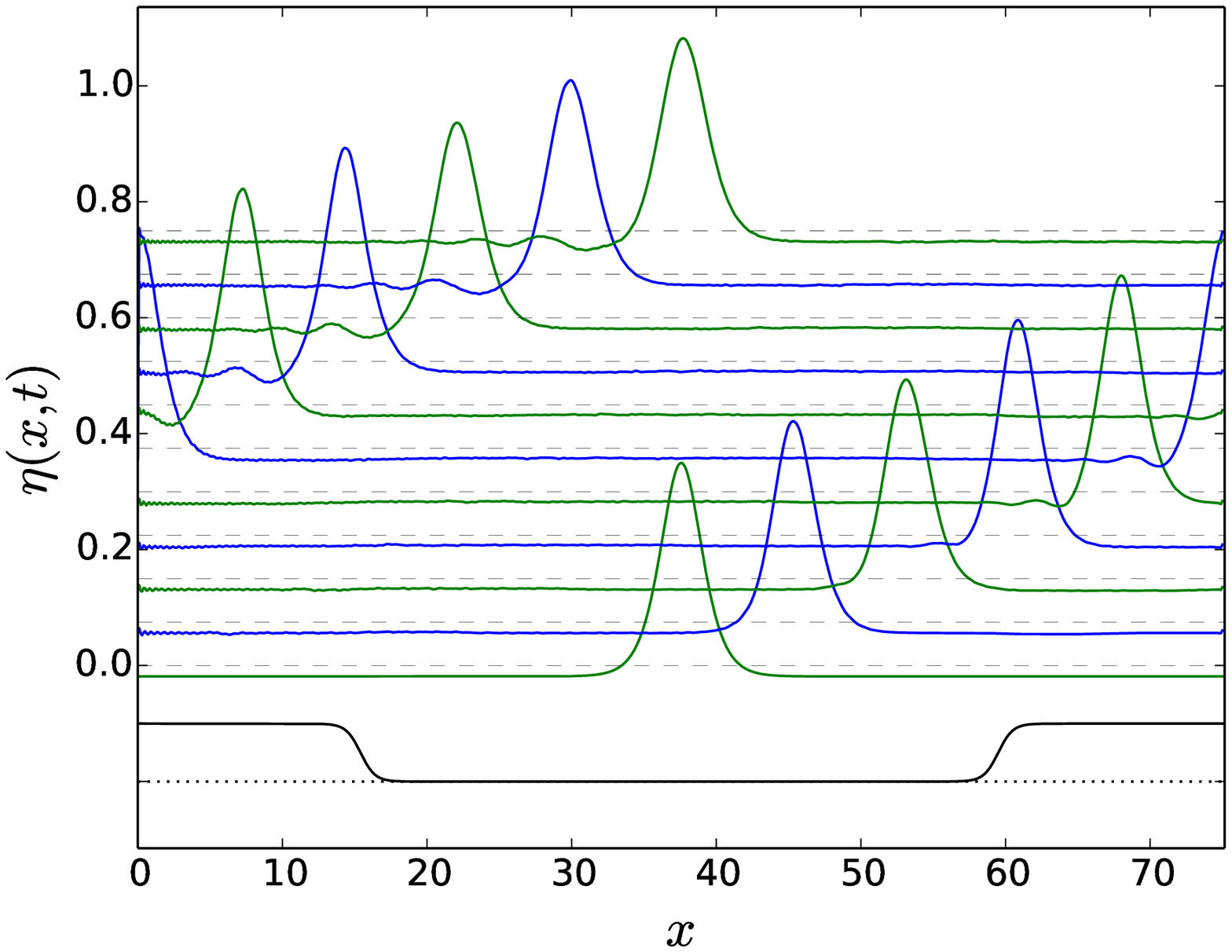}}
\vspace{-2mm}
\caption{Time evolution, according to eq.~(\ref{etaabd}), of the cnoidal wave for the bottom function in the form of an extended valley. The $x$ interval is equal to the wavelength of the cnoidal wave with $m$=1-10$^{-16}$.} 
 \label{m16000}
\end{figure}

\begin{figure}[tbh]  
\resizebox{0.99\columnwidth}{!}{\includegraphics{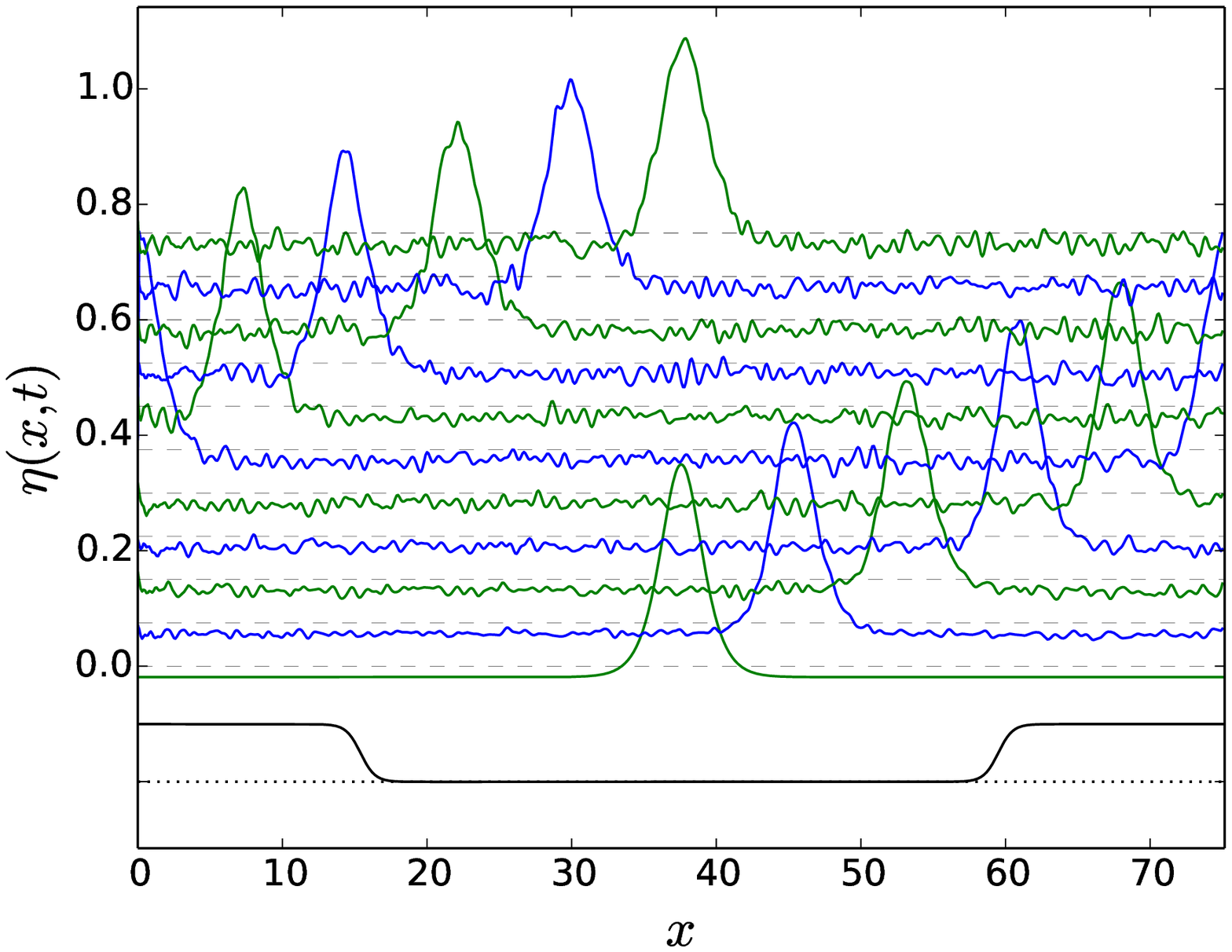}}
\vspace{-2mm}
\caption{The same as in Fig.~\ref{m16000} but with a moderate amplitude of stochastic force, $\gamma=0.001$} 
 \label{m16001}
\end{figure}

\begin{figure}[tbh]  
\resizebox{0.99\columnwidth}{!}{\includegraphics{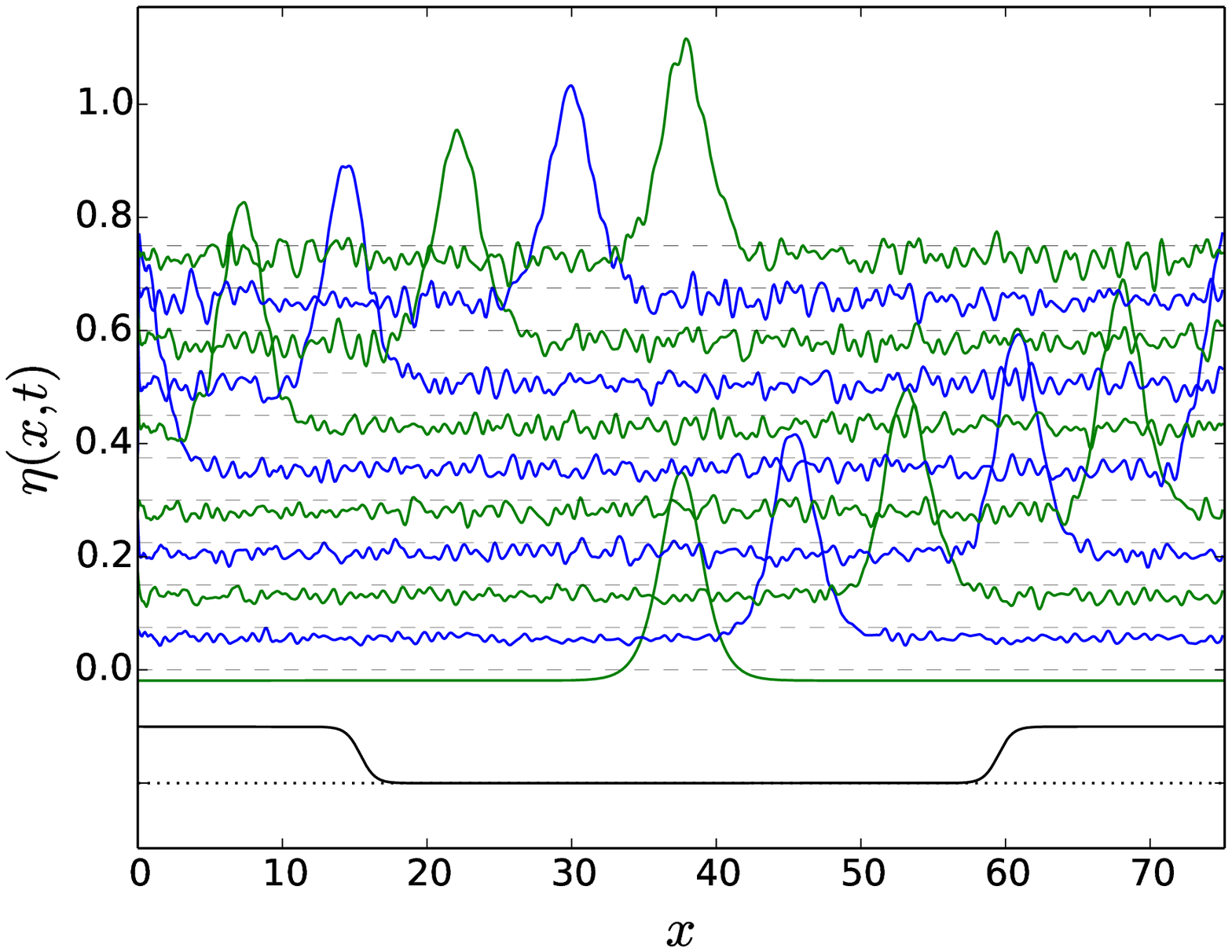}}
\vspace{-2mm}
\caption{The same as in Fig.~\ref{m16001} but with a larger amplitude of stochastic force, $\gamma=0.0015$.} 
 \label{m16002}
\end{figure}

\section{Conclusions} \label{wnioski}

The main conclusions obtained from our numerical simulations of time evolution of KdV-type waves with respect to second order equations with bottom terms are the following:
\begin{itemize}
\item Both solitary and cnoidal solutions of KdV equations are extremely robust structures for many possible distortions. In previous studies \cite{KRR,KRI,KRI2} we 
showed the resistance of these waves to second order terms in extended KdV equation, including terms from an uneven bottom.
\item In this paper we showed that an inclusion of a stochastic force into second order KdV-type equation does not disturb much the shape of the main wave even for large amplitude of the noise, although the secondary wave structures can be completely obscured by the noise. It seems, however, that the main wave is the most robust for solitary waves (which is a limitnig case of cnoidal waves  when $m$ tends to 1). 
That robustness with respect to stochastic noise diminishes when parameter $m$ decreases below~1.
\item
Finite element method, though sufficient for numerical study of stochastic KdV equation in \cite{DebP} is not  so well suited for the higher order KdV type equation, and particularly  less satisfactory when the bottom is not flat. For KdV equation considered in a moving frame (\ref{kdv}) the wave motion is slow and important time evolution can be calculated using not very long space and time intervals. This property allowed the authors of \cite{DebP} to use relatively low number $N=200$ of the mesh size to obtain  relevant results. Consequently, since KdV is a third order differential equation, the size of Jacobian matrix used in numerical scheme, $(3N\times 3N)$ is still low and allows for fast calculations. 
Second order ("extended") KdV equation (\ref{etaab}), which is a differential equation of fifth order, can be studied both in a moving reference frame and in a fixed frame. In the first case the size of the Jacobian increases to $(5N\times 5N)$ and when $N$ is of the same order the results can still be obtained in reasonable computing time. The
equation taking into account bottom variation (\ref{etaabd}), however, has to be solved in the fixed frame. Then, since waves move much faster, in order to obtain a good description of the wave evolution, substantially longer space intervals have to be used.
For a resolution of fine structures of the wave relatively dense mesh has to be applied so $N$ is usually an order of magnitude larger than in the case of moving reference frame.  Then computer time for inversion of the Jacobian becomes very large and detail calculations are not practical. In these cases the finite difference method used in \cite{KRR,KRI}, adapted for stochastic equation, proves to be more efficient. 
\end{itemize}

\end{document}